\documentstyle[prl,aps,preprint]{revtex}
\def\scr#1{{\cal #1}}
\def\dk{\frac{d^4k}{(2 \pi)^4}}

\def\ket#1{| #1 \rangle}

\def\al{\alpha}

\def\De{\Delta}

\def\roughly#1{\mathrel{\raise.3ex\hbox{$#1$\kern-.75em%
    \lower1ex\hbox{$\sim$}}}}
\def\lsim{\roughly<}

\begin{document}
\setcounter{page}{0}
\def\footnoterule{\kern-3pt \hrule width\hsize \kern3pt}
\tighten
\preprint{\vbox{
\hbox{MIT-CTP-25XX}
\hbox{hep-th/9511365}}}
\title{CHIRAL PERTURBATION THEORY ANALYSIS OF BARYON TEMPERATURE
MASS SHIFTS\thanks
{This work is supported in part by funds provided by the U.S.
Department of Energy (D.O.E.) under cooperative
research agreement \#DF-FC02-94ER40818.}}

\author{Paulo F. Bedaque\footnote{Email address: {\tt BEDAQUE@ctpa04.mit.edu}}}
\smallskip
\address{Center for Theoretical Physics \\
Massachusetts Institute of Technology \\
Cambridge, Massachusetts 02139 \\
{~}}

\date{October 1995}
\maketitle

\thispagestyle{empty}

\begin{abstract}

We compute the finite temperature pole mass shifts of the octet and
decuplet baryons using heavy baryon chiral perturbation theory and
the $1/N_c$ expansion, where $N_c$ is the number of QCD colors.
We consider the temperatures
of the order of the pion mass $m_{\pi}$, and expand truncate the
chiral and $1/N_c$ expansions assuming that $m_\pi \sim 1/N_c$.
There are three scales in the problem:
the temperature $T$, the pion mass $m_\pi$, and the octet--decuplet
mass difference.
Therefore, the result is not simply a power series in $T$.

We find that the nucleon and $\De$ temperature mass shifts are opposite
in sign, and that their mass difference changes by $20 \%$
in the temperature range $90\ {\rm MeV} < T < 130\ {\rm MeV}$,
that is the range where the freeze out in relativistic heavy ion collisions
is expected to occur.

We argue that our results are insensitive to the neglect of $1/N_c$-
supressed effects;
the main purpose of the $1/N_c$ expansion in this work is to justify
our treatment of the decuplet states.

\end{abstract}

\vspace*{\fill}
\begin{center}
Submitted to: {\it Physical Review D}
\end{center}

\pacs{xxxxxx}
\section{Introduction}

In this paper we compute the temperature-dependence of the pole mass
of the $J=\frac{1}{2}$ and $J=\frac{3}{2}$ baryons using effective
lagrangians and large $N_c$ methods.
There are two main motivations for this work.
The first comes from relativistic heavy ion collisions.
These are very complex phenomena, and the extraction of physically
interesting quantities from the experimentally accessible information
necessarily involves a good deal of modelling.
Most likely, a number of measurable quantities will have to be correlated
to constrain the ones of physical interest.
Clearly, an understanding of the parameters of these models within
controlled approximations will simplify this process.
The dependence of particle properties on the temperature is one such
parameter.
The second motivation is more theoretical.
The finite temperature properties of hadrons provide clues to the
mechanism of confinement/chiral symmetry breaking which is expected
to occur at $T \simeq 200\ {\rm MeV}$.
Below this critical temperature, we can use effective lagrangian
methods to gain some information about the finite-temperature dependence
of QCD quantities.

Chiral perturbation theory ($\chi$PT) is the effective low-energy
theory of hadrons and gives rise to a systematic expansion in powers
of the light current quark masses $m_{u,d,s} / \Lambda$ and kinematic
scales $E / \Lambda$, where $\Lambda \sim 1\ {\rm GeV}$ is the
chiral expansion scale.
All the information contained in QCD is encapsulated in a few
phenomenological constants, that have to be determined by experiment.
An essential result is that at a given order in the expansion
one has to compute only diagrams with a finite number of loops depending
on a finite number of couplings~\cite{ref:pioneers}.
Tree diagrams correspond to current algebra results, of which $\chi$PT
can be viewed as a systematic extension to all orders.
Formulated originally for the Goldstone bosons $\pi$, $K$, and $\eta$,
it was later extended to include the low-lying
baryons~\cite{ref:baryonbegin}.
A non-trivial aspect of this extension is the presence of a new scale,
the baryon mass, that does not vanish in the chiral limit.
This appears to spoil the scaling arguments used to get the relation
between number of loops and powers of quark mass (or momentum).
An elegant solution to this problem is given by heavy baryon chiral
perturbation theory (HB$\chi$PT)~\cite{ref:heavybaryon},
which treats the baryon as a static particle.
Corrections suppressed by powers of the baryon mass appear as
higher-order vertices in the effective lagrangian.
Since the baryon mass appears only in vertices, the chiral counting
manifestly goes through and the relation between the number of loops
and powers of quark mass is recovered.

One difficulty with HB$\chi$PT is the large number of phenomenological
constants appearing in the effective lagrangian.
Another one is the presence of $\Delta$ states close to the nucleons
($M_\Delta - M_N \simeq 300$ MeV).
There are different prescriptions for dealing with the decuplet states:
some authors include them as explicit degrees of freedom (treating the
the decuplet--octet mass difference either perturbatively or to all
orders), while others keep only the nucleon fields and consider only
processes at energies low enough to avoid exciting the decuplet states.
The $1/N_c$ expansion (where $N_c$ is the number of QCD colors) alleviates
both of these problems~\cite{ref:largeN}.
It relates the value of some phenomenological constants, allowing one to
go further in the chiral expansion.
Also, in the $1/N_c$ expansion there is a whole tower of degenerate
baryon states with spins $\frac{1}{2}, \frac{3}{2}, \ldots$ with
mass differences between low-lying spin multiplets suppressed by $1/N_c$.
The $1/N_c$ expansion therefore forces us to include the decuplet
as explicit degrees of freedom, and (after deciding how to book $1/N_c$
corrections compared to chiral corrections) the treatment of the
decuplet--octet splitting is determined.
Another appealing aspect of the $1/N_c$ expansion is that the intuitive
and phenomenologically successfull static quark model arises as its leading
term ~\cite{ref:absbegg}.
HB$\chi$PT together with the $1/N_c$ expansion has been used to
compute a number of quantities sucessfully~\cite{ref:ex}.
In particular, a discussion of the zero temperature baryon
mass differences at next-to-leading order will be reported
elsewhere~\cite{ref:we}.

At temperatures $T$ well below the critical temperature
$T_c \sim 200$ MeV, we expect a description of a QCD plasma in terms
of hadronic degrees of freedom to be justified.
Since most particles will have kinetic energies of order $T$ or
smaller, their interactions will be soft and we expect that $\chi$PT
will give an accurate description of the dynamics.
Particles with high energy or large masses are exponentially
suppressed by Boltzmann factors and can be safely ignored.
For example, Ref.~\cite{ref:leut_on_cc} studied the thermodynamics
of a gas of pions to compute the temperature dependence of the
chiral condensate.

The properties of baryons in a thermal plasma are affected by the
background of thermal pions.
In this paper we compute the leading order temperature dependence of
the pole mass of octet and decuplet baryons  using
HB$\chi$PT and the $1/N_c$ expansion.

This paper is organized as follows.
In section II we explain the power counting that selects the class
of diagrams contributing to the leading order temperature baryon
pole mass dependence.
In Section III the results are presented and discussed.
The chiral limit is briefly considered in Section IV where a few
comparisons with previous results are made.
Finally, in Section V the main results are summarised and comments
on the range of validity of our approach are presented.

\section{Effective lagrangian and power counting}
The calculation will involve the real time formalism (RTF) for finite
temperature field theory and the formalism presented in~\cite{ref:lmr}
for large $N_c$ heavy baryon chiral perturbation theory.
Here we confine ourselves to defining the notation used, leaving the
details to the references.

The complications caused by the presence
of a matrix structure in the propagators in the RTF~\cite{ref:rtf} is
offset by the fact that it is easy to sepatate the temperature-dependent
contributions from the zero temperature part.
The propagator for the (static) baryon with velocity $v^\mu$
(taken to be the same as the plasma velocity) is
\begin{mathletters}%
\label{eq:1all}
\begin{eqnarray}
   S_{11}(x)&=& \int \dk e^{-ikx}
   ~\left[{1\over  v\cdot k+i\epsilon} +
      2\pi i n_F(M+v\cdot k)\delta(v\cdot k)\right]\ ,
   \label{eq1:a}\\
   S_{22}(x)&=& \int \dk e^{-ikx}
   ~\left[-{1\over v\cdot k-i\epsilon}
      + 2\pi i n_F(M+v\cdot k)\delta(v\cdot k)\right]\ ,
   \label{eq1:b}\\
   S_{12}(x)&=& \int \dk e^{-ikx}
   ~\left[2\pi i (n_F(M+v\cdot k)
      - \theta(-M-v\cdot k))\delta(v\cdot k)\right]\ ,
   \label{eq1:c}\\
   S_{21}(x)&=& \int \dk e^{-ikx}
   ~\left[2\pi i (n_F(M+v\cdot k)
       - \theta(M+v\cdot k))\delta(v\cdot k)\right]\ ,
   \label{eq1:d}
\end{eqnarray}
\end{mathletters}
and the meson propagator is
\begin{mathletters}%
\label{eq:2all}
\begin{eqnarray}
D_{11}(x)&=& \int \dk e^{-ikx}
   ~\left[{1\over k^2-m^2+i\epsilon}
       - 2\pi i n(v\cdot k)\delta(k^2-m^2)\right]\ ,
    \label{eq2:a}\\
            D_{22}(x)&=& \int \dk e^{-ikx}
    ~\left[-{1\over k^2-m^2-i\epsilon}
           - 2\pi i n(v\cdot k)\delta(k^2-m^2)\right]\ ,
     \label{eq2:b}\\
     D_{12}(x)&=& -\int \dk e^{-ikx}
     ~\left[ 2\pi i (n(v\cdot k) + \theta(-v\cdot k))\delta(k^2-m^2)\right]\ ,
     \label{eq2:c}\\
               D_{21}(x)&=& -\int \dk e^{-ikx}
        ~\left[2\pi i (n(v\cdot k) + \theta(v\cdot k))\delta(k^2-m^2)\right]\ ,
           \label{eq2:d}
\end{eqnarray}
\end{mathletters}
where $n(x)$ ($n_F(x)$) is the bosonic (fermionic) distribution function
\begin{equation}
n(x) = {1\over e^{|x|}-1},
\qquad n_F(x)={1\over e^{|x|}+1}.
\end{equation}
The indices $1,2$ refer to the the matrix structure of the RTF propagator
needed to describe the expectation value of operators in a mixed state.
Note the presence of the large baryon mass $M$ in the distribution function
$n_F$.
They appear because, in  HB$\chi$PT, the momentum in the propagator%
{}~(\ref{eq:1all}) is the residual momentum $k = p - Mv$ where $p$ is the
physical momentum~\cite{ref:heavybaryon}.
Thus, in the range of temperatures considered here, the
temperature-dependent part of the baryon propagator is suppressed
by a tiny Boltzmann factor and can be safely dropped.
To a lesser extent, the same is true for $K$ and $\eta$ mesons:
since $n(m_{\pi}/T) / n(m_K/T) \sim 10$ at $T=200$ MeV, we will count
thermal kaon and eta loops as addionally suppressed in the chiral
counting.
In the present leading order
calculation this means they will also be dropped.

In the large-$N_c$ world, the low lying baryons form an infinite tower
of flavor multiplets with spin $\frac{1}{2}, \frac{3}{2}, \ldots$.
We follow the formalism and notation of Refs.~\cite{ref:lmr,ref:carone}
where they are all treated simultaneously by defining a field acting on
a spin--flavor Fock space
\begin{equation}
|B)=B^{a_1 \al_1 \cdots a_N \al_N}
\hat a^\dagger_{a_1 \al_1 } \cdots
\hat a^\dagger_{a_N \al_N } \ket{0},
\end{equation}
where $a_1, \ldots, a_N = 1, \ldots, N_F$ are flavor indices and
$\al_1, \ldots, \al_N = \uparrow, \downarrow$ are spin indices.
The operators $\hat a^\dagger$ and $\hat a$ are (bosonic) creation and
annihilation operators.
The effective lagrangian then has the standard form
\begin{equation}
L=(B| \{i v\cdot D\}+ \mu \{\sigma_\mu\}\{\sigma^\mu\} - \{\scr M\}+
             g \{A_\mu \sigma^\mu\}+ \dots |B)
\end{equation}
where the braces around a generic operator stand for
\begin{equation}
\{W\}=W^{a \al}{}_{b \beta}\hat a^\dagger_{a \al }\hat a^{b \beta }.
\end{equation}
$\cal M$ is built from the quark mass matrix $M$
\begin{equation}
{\cal M} = {1\over 2} ( \xi M \xi + \xi^\dagger M \xi^\dagger),
\end{equation}
and
\begin{eqnarray}%
D |B)&\equiv&(\partial - i \{V\})|B),\\
V_\mu& \equiv& \frac i2\left(\xi \partial_\mu \xi^\dagger
+ \xi^\dagger \partial_\mu \xi\right), \qquad\\
A_\mu &\equiv& \frac i2\left(\xi \partial_\mu \xi^\dagger
-\xi^\dagger \partial_\mu \xi\right),
\end{eqnarray}%
where $\xi(x)$ parametrizes the Goldstone bosons
\begin{equation}
\xi(x) = e^{i\Pi(x) / f},
\end{equation}
and $f\simeq 93$ MeV is the pion decay constant.

The main result obtained in Ref.~\cite{ref:lmr} is that the coefficient
of an $n$-body operator in the effective lagrangian above has a coefficient
which is $\lsim 1/N_c^{n-1}$ for large $N_c$.
This comes about because, at the QCD level, an exchange of $n-1$ gluons
is necessary to produce a term in the effective lagrangian involving $n$
quarks.
This is true even when the number of light quark flavors $N_F$ is taken
to be of order $N_c$, since  this rule does not rely in the suppresion
of quark loops~\cite{ref:nf}.
For $N_F \ge 3$, the size of the flavor representations
of the low lying baryons with given $J\sim 1$ grow with $N$
and it is not obvious which of those states
should be identified with the states of the real  $N=3$ world.
Our procedure will be to compute the mass shifts keeping all terms
that are of the order
desired in {\it any} state in the large $N$ flavor multiplet. Only
then those terms are evaluated
with $N=N_F=3$.

Because we take $N_F \sim N_c$, a decision has to be made about how
to extrapolate the quark mass matrix to arbitrary number of flavors.
We choose the most general form
\begin{equation}
M = \left (
\begin{array}{c c c} m_u 1_{N_u} & 0 & 0 \\
                       0 & m_d 1_{N_d} & 0 \\
                       0 & 0 & m_s 1_{N_s} \\
\end{array}
\right ),
\end{equation}
and keep all terms that are of the desired order in any extrapolation
in which $N_u + N_d + N_s = N_F$.

Now we can establish the rules for power counting.
We take $1/N$ to be of the same order as a strange quark mass $m_s$.
The light quark masses $m_{u,d}$ are counted as $\sim 1/N^2$.
As mentioned above, kaon and eta thermal loops are suppressed by
large Boltzman factors compared to pion thermal loops, and are therefore
negligible in this expansion.

This suppression however, does not become
exact in the chiral or large $N$ limit. To account for this numerical fact
we take the practical attitude of counting those loops as
additionally suppressed by a factor $\sim 1/N^2$.
Summarizing,
\begin{equation}%
\frac 1N  \sim m_s  \sim \epsilon,
\qquad \hat m = \frac 12 (m_u+m_d)  \sim K,\eta\ {\rm thermal\ loop} \sim
\ \epsilon^2.
\end{equation}

Let us now look at the graphs giving the leading temperature dependent
contributions to the baryon mass.
First, consider the graph in Fig.~1.
The tick on the meson propagator denotes the temperature dependent
part of the propagator.
This graph contains two one-body vertices, and two
powers of $f$ in the denominator.
Since $f \sim \sqrt{N_c}$ and the matrix element of any $n-$body
operator is at most $\sim N_c^n$, using dimensional analysis we can
see that this graph is $\sim N m_{\pi}^3 I(m_{\pi}/T)$, where
$I(x)$ is a some dimensionless function.
For $T \sim m_{\pi}$ we have $I(m_{\pi}/T) \sim 1$, so this graph is
booked as $N \epsilon^3$.
An explicit calculation shows, however, that $I(m_{\pi}/T)$ vanishes.

The graphs in Fig. 2, with an arbitrary number $r$ of mass insertions
of the form $\mu \{ \sigma^i\}\{ \sigma^i\}$  are of the order
$ N^{1-r} m_{\pi}^{3-r} \sim N \epsilon^3$
(since $\mu \sim 1/N$).
The sum of these graphs therefore gives the leading contribution to the
temperature mass shift.

Note that na\"\i vely, the diagram in Fig.~2 with one insertion of the
operator $\{ \cal M \}$ is of order $N_c^2 m_s m_{\pi}^2$.
This contradicts the fact that baryon masses are proportional to $N$.
When wave function renormalization terms are included however,
there is a cancellation that decreases this result by two powers of
$N_c$, making this contribution not only consistent with large $N_c$
expectations, but actually negligible at leading order in our expansion.
We assume that similar cancellations (necessary for the
consistency of large $N_c$ QCD) occur so that no contribution grows
faster than $\sim N_c$.
This has been checked in various non trivial cases ~\cite{ref:we}.

Using similar arguments, one can check that other graphs including
$1/N_c$ suppressed or higher derivative vertices, as well as light
quark mass insertions and higher loop graphs are also of higher order.
Thus, the leading contribution for the temperature
dependent part of the pole mass is given by the simple one loop graph
with the baryon propagator resummed to include the (zero temperature)
octet-decuplet mass splitting.
Since this graph is to be evaluated at the tree level mass shell
(including the effects of the term $\mu \{\sigma^i \} \{\sigma^i \}$),
the external momentum will cancel against the mass
in the propagator if both the internal and external particles, have
the same spin.
When this happens the graph with the resummed propagator is exactly
equal to the graph in Fig.~1, which vanishes.
Thus, the only intermediate states that contribute to the octet temperature
mass shifts are decuplet intermediate states, and {\it vice-versa}.

\section{Results}
The calculation of the graph in Fig.~2 with the resummed propagator is
straighforward.
Keeping only the temperature dependent part we have
\begin{eqnarray}
\label{eq:result}
\delta_T M &=&\sum_{A=1}^{3}\left( \frac{g}{\sqrt{2} f}\right)^2
           \int\dk k^i k^j
           \frac{1}{v\cdot k \mp \mu + i \epsilon}\, ( -2\pi i)\, n(v\cdot k)
           \,\delta (k^2-m_{\pi}^2)\,
           \{ T^A \sigma^i\} {\cal P}\{ T^A \sigma^j\}
\nonumber\\
&=&\Biggl[\pm \frac{g^2 \mu m_{\pi}^2}{ 12 \pi ^2 f^2}
\int_0^\infty dx
{x^4\over \sqrt{x^2+1}}\quad \frac{1}{x^2+1 - {\mu^2}/ {m_{\pi}^2}}
\quad \frac{1}{e^{\sqrt{x^2+1}\, {m_{\pi} / T}}-1}
\nonumber\\
& & \quad -\, i{g^2\over 24 \pi f^2} (\mu^2 - m_{\pi}^2)^{\frac 32}
\frac{1}{e^{\mu / T}-1} \Biggr]\, {\cal X}\ ,
\end{eqnarray}
where $T^A$ are $SU_F(3)$ generators and ${\cal P}$ is a projection
operator
\begin{equation}
{\cal P}_8 \equiv {15-\{ \sigma^i\}\{ \sigma^j\} \over 12},
\qquad {\cal P}_{10} \equiv {\{ \sigma^i\}\{ \sigma^j\}-3 \over 12}.
\end{equation}
Here ${\cal P}_8$ is the projection operator for octet masses, which
projects onto decuplet intermediate states;
{\it vice versa} for the decuplet masses.
The upper(lower) sign in (~\ref{eq:result}) applies to the octet
(decuplet) and ${\cal X}$ is the operator
\begin{eqnarray}
{\cal X}_8 & = & 15 - {34\over 3} \{ S\}
+ {8\over 3} \{ S \sigma^i\}\{\sigma^i\}+
\frac 13\{S\}\{S\},
\\
{\cal X}_{10} & = & 9 +{34\over 3} \{ S\} -
{8\over 3} \{S \sigma^i\}\{\sigma^i\}-
\frac 13\{S\}\{S\},
\end{eqnarray}
Here,
\begin{equation}
S = \left (
\begin{array}{c c c c} 0 & 0 & 0 \\
                       0 & 0 & 0 \\
                       0 & 0 & 1 \\
\end{array}
\right )
\end{equation}
is the strangeness matrix in flavor space.
The expectation value of $X$ in different states is
\begin{eqnarray}
\label{eq:x}
{\cal X}_N&=15,\nonumber \\
{\cal X}_\Sigma&=\frac 43,\nonumber \\
{\cal X}_\Lambda&=12,\nonumber \\
{\cal X}_\Xi&={13\over 3},\nonumber \\
{\cal X}_\Delta&= 9,\nonumber \\
{\cal X}_{\Sigma^\ast}&={20\over 3},\nonumber \\
{\cal X}_{\Xi^\ast}&={13\over 3},\nonumber \\
{\cal X}_\Omega&=0.\nonumber \\
\end{eqnarray}
The zero value for  ${\cal X}_\Omega$ reflects the fact that
the $\Omega$ interactions with the background pions involve the
annihilation of two quarks into
gluons that then turn into the quark making up the pion, which is a $1/N$
suppressed process.

The integral in (~\ref{eq:result}) is to be understood in the principal
value sense.
The coupling constant $g$ is related to the usual octet axial couplings
$D$ and $F$, as well as the axial couplings to the decuplet states.
In the large-$N_c$ and $SU_F(3)$ limits we can identify as
$g = 3(D+F)/5 \simeq 0.8$, which is also the value obtained by
performing a best fit at leading order.
In Fig.~3 we show the dependence of the real part of the nucleon pole
mass as a function of temperature.
(The temperature dependence dependence for other octet baryons is a
multiple of the same graph.)
A striking feature is that it is very small until $T\simeq 80$~MeV, and
then varies rapidly.

The shift in the mass of the $\De$ is opposite to the nucleon, so that the
mass splitting between then decreases rapidly in the range $90~ {\rm Mev}<
T< 130 ~{\rm Mev}$. This fact may be used to predict relative yields of
different baryon species in central rapidity region of relativistic
heavy ion collisions. Assuming that the number baryons of a given kind
is given, in
first approximation, by the Fermi distribution at the freeze out temperature,
the relative yields of two baryons will be a function of their mass splittings.
As a correction to this model, it is natural to use the temperature
dependent masses in the
Fermi distribution (the precise sense in which this gives the leading
correction will be the subject of a future publication). It may be possible
then to test our predictions by measuring ratios of the number of different
baryons species. It is not clear however that this simple freeze out models
are accurate within $20 \%$ so that the relatively small mass shifts can be
observed.

The imaginary part of the pole is shown in Fig.~4.
Note that this width can be caused not only by the decay of the incoming
particle, but by other processes without counterpart at zero
temperature, such as the absorbtion of a pion from the thermal
background~\cite{ref:weldon};
even stable particles can have a finite width at nonzero temperature.
For the temperatures where we expect $\chi$PT to be reliable, say,
$T < 200$ MeV, the width of the nucleons is small compared to their masses,
and they therefore behave as sharp, well-defined states.

A hint of deconfinement is found when the residue of the baryon poles
$Z^{-1}$ is calculated.
This computation is very similar to the one presented above,
the main difference being that now that both the octet and the
decuplet contribute for all states.
For example, the nucleon residue $Z^{-1}$ changes by an ammount
\begin{equation}
\label{eq:z}
\delta_T Z^{-1}= {27\over  2} H(0) + 15 H(-\mu),
\end{equation}
where
\begin{equation}
H(x)={g^2\over 12 \pi^2 f^2} \int_0^\infty dk\,
 {k^4 n(\omega_k)\over \omega_k}
     {x^2+\omega_k^2\over (x^2-\omega^2+i \epsilon)^2}.
\end{equation}
At $T=200$ MeV, for instance, the residue is reduced by $15\%$ in relation
to the zero temperature value, and decreases very fast with higher
temperatures.
This is consistent with the picture that the baryon poles disappear at
high temperature.

\section{Chiral Limit}
It is customary to consider the chiral limit where the quark masses
vanish when using chiral models at finite tempertaure.
This limit has the theoretically pleasing feature that the only mass
scales in the problem are the temperature $T$ and the chiral expansion
scale, and all quantities have a power series expansion in $T$.
We note here, however, that this limit is usefull when $T \gg m_{\pi}$,
which is not of great phenomenological interest given the fact that
the critical temperature is numerically close to the pion mass.

Nonetheless, it is worth considering this limit as a consistency
check.
In the chiral limit $m_{u,d} \rightarrow 0$, our main result
Eq.~(\ref{eq:result}) is
\begin{eqnarray}
\label{eq:chirallimit}
\delta_T M &=& \Biggl[ \pm \frac{g^2 \mu T^2}{\pi ^2 f^2}
\int_0^\infty dx\,
\frac{x^3}{x^2 - {\mu^2 / T^2}}\,
\frac{1}{e^x-1}
\nonumber\\
& & \quad -\, i \frac{g^2}{24 \pi f^2} \mu^3
{1\over e^{\mu / T}-1}\Biggr] \,{\cal X}.
\end{eqnarray}
This is not a power series in $T$ even in the chiral limit because of
the presence of the additional mass scale $\mu$ (the decuplet--octet
mass difference).
It becomes a power series in $T$ in the limit $\mu \rightarrow 0$,
in which case the baryon mass shift vanishes.
It also becomes a power series when $\mu \rightarrow \infty$,
corresponding to decoupling the decuplet states;
This is the limit which corresponds to ``pure'' HB$\chi$PT.
In this limit, we obtain
\begin{equation}
\label{eq:T4}
\delta_T M=\mp {g^2 T^4\over 72 f^2 \mu}+ {\cal O}(1/\mu^2).
\end{equation}
This is consistent with a general theorem that states that there is
no contribution at order $T^2$ in the chiral limit.
This is also in a agreement with the calculation of
Ref.~\cite{ref:leut_on_baryons}) which used a relativistic effective
lagrangian for the baryons and concluded that there are no
contributions to the baryon mass at order $T^2$ in the chiral limit.

\section{Conclusion}
We have computed the temperature dependence of the pole mass of baryons
in leading order in HB$\chi$PT and the $1/N_c$ expansion for temperatures
$T \lsim m_{\pi}$.
The full dependence on the temperature and on the octet--decuplet
mass difference is included, as dictated by large-$N_c$ and chiral
power counting.
The shifts in mass were of the order of $20\%$, at $T\simeq 150$ MeV, and
the changes in width are about the same size.

The range of temperatures where our results are applicable are limited
by both the chiral expansion (which breaks down at energies of the order
of QCD scales, like the rho mass) and by the existence of heavier
resonances.
The effect of virtual loops of these heavy particles
at zero temperature are included in the value of the phenomenological
coefficients of the effective lagrangian.
Although such particles are also present in the thermal plasma, they
are suppressed by large Boltzmann factors at low temperatures, but
at higher temperatures, they will become important.

It is important to realize that the large-$N_c$ expansion entered in
the present calculation in a rather non-essencial way.
As long as we decide to include the decuplet as a propagating field,
consider the octet-decuplet splitting of the order $\sim m_s$, and use
a value for the octet-decuplet coupling constant consistent with its
large-$N_c$ value, our results are essentially unaltered.
Only a numerically very small term was dropped (since it was $1/N_c$
suppressed) in arriving at (~\ref{eq:result}).

\section{Acknowledgements}
I would like to thank Markus Luty for extensive discussions on this
and related subjects.

\vfill
\eject
\centerline{FIGURE CAPTIONS}

\bigskip
\bigskip
Fig. 1- Contribution of order $N m_{\pi}^3$ to the baryon mass.
Full lines denote baryons, dashed lines with a  tick stand
for the temperature dependent piece of the meson propagator.

Fig. 2- Leading non vanishing contribution to the temperature dependent
baryon mass. The crosses stand for mass insertions.

Fig. 3- Real part of the mass pole as a function of temperature.

Fig. 4- Imaginary part of the mass pole as a function of temperature.

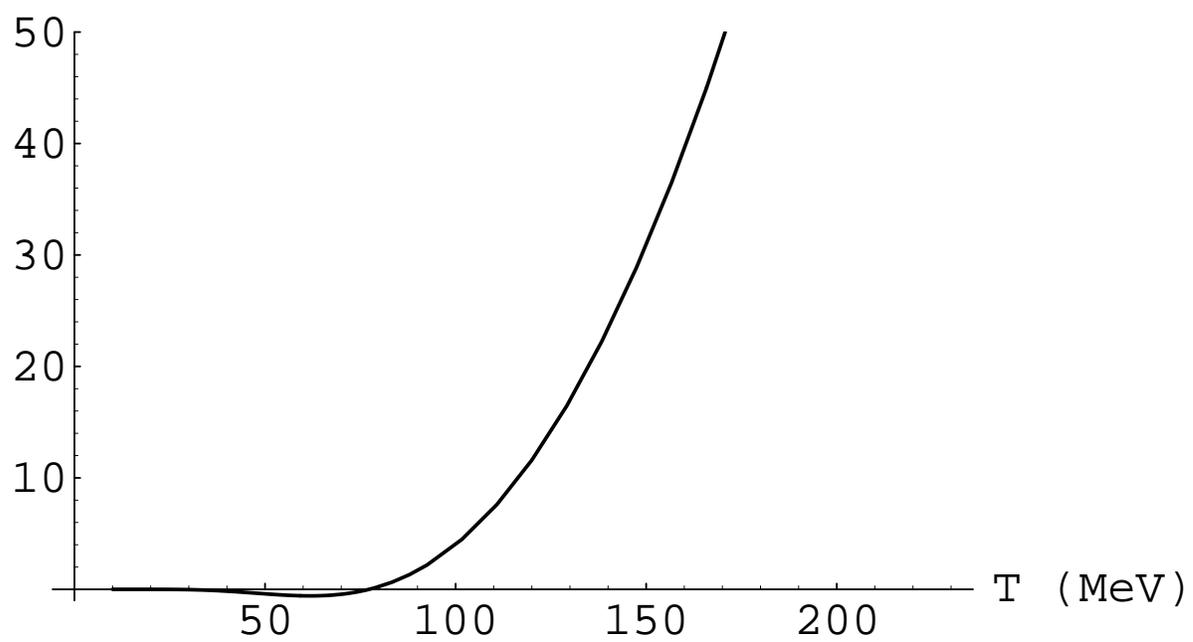
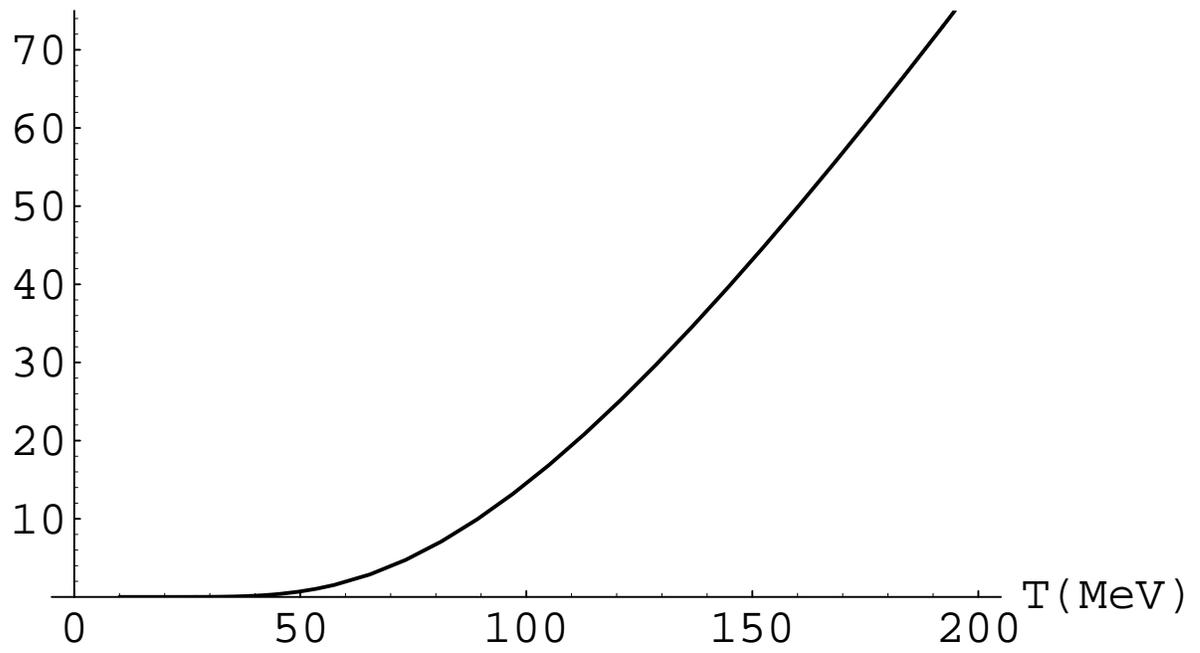
\end{document}